\documentclass[reprint,showkeys,superscriptaddress,amsmath,amssymb,aps,prl]{revtex4-1}

\usepackage{graphicx}
\usepackage{hyperref}
\usepackage{url}
\usepackage{epstopdf}
\usepackage{color}
\definecolor{orange}{RGB}{255,165,0}
\usepackage[]{siunitx}

\begin{document}

\title{Fermi-surface reconstruction at the metamagnetic high-field transition in uranium mononitride}

\author{S. Hamann}
\affiliation{Hochfeld-Magnetlabor Dresden (HLD-EMFL) and W\"urzburg-Dresden Cluster of Excellence ct.qmat, Helmholtz-Zentrum Dresden-Rossendorf (HZDR), 01328 Dresden, Germany}
\affiliation{Max Planck Institute for Chemical Physics of Solids, 01187 Dresden, Germany}
\author{T. Förster}
\affiliation{Hochfeld-Magnetlabor Dresden (HLD-EMFL) and W\"urzburg-Dresden Cluster of Excellence ct.qmat, Helmholtz-Zentrum Dresden-Rossendorf (HZDR), 01328 Dresden, Germany}
\author{D.~I. Gorbunov}
\affiliation{Hochfeld-Magnetlabor Dresden (HLD-EMFL) and W\"urzburg-Dresden Cluster of Excellence ct.qmat, Helmholtz-Zentrum Dresden-Rossendorf (HZDR), 01328 Dresden, Germany}
\author{M. König}
\affiliation{Max Planck Institute for Chemical Physics of Solids, 01187 Dresden, Germany}
\author{M. Uhlarz}
\affiliation{Hochfeld-Magnetlabor Dresden (HLD-EMFL) and W\"urzburg-Dresden Cluster of Excellence ct.qmat, Helmholtz-Zentrum Dresden-Rossendorf (HZDR), 01328 Dresden, Germany}
\author{J. Wosnitza}
\affiliation{Hochfeld-Magnetlabor Dresden (HLD-EMFL) and W\"urzburg-Dresden Cluster of Excellence ct.qmat, Helmholtz-Zentrum Dresden-Rossendorf (HZDR), 01328 Dresden, Germany}
\affiliation{Institut f\"ur Festk\"orper- und Materialphysik, TU Dresden, 01062 Dresden, Germany}
\author{T. Helm}
\email[]{t.helm@hzdr.de}
\affiliation{Hochfeld-Magnetlabor Dresden (HLD-EMFL) and W\"urzburg-Dresden Cluster of Excellence ct.qmat, Helmholtz-Zentrum Dresden-Rossendorf (HZDR), 01328 Dresden, Germany}
\affiliation{Max Planck Institute for Chemical Physics of Solids, 01187 Dresden, Germany}

\date{\today}

\begin{abstract}
We report on the electronic and thermodynamic properties of the antiferromagnetic metal uranium mononitride with a Néel temperature $T_N\approx 53\,$K. The fabrication of microstructures from single crystals enables us to study the low-temperature metamagnetic transition at approximately $58\,$T by high-precision magnetotransport, Hall-effect, and magnetic-torque measurements. We confirm the evolution of the high-field transition from a broad and complex behavior to a sharp first-order-like step, associated with a spin flop at low temperature. In the high-field state, the magnetic contribution to the temperature dependence of the resistivity is suppressed completely. It evolves into an almost quadratic dependence at low temperatures indicative of a metallic character. Our detailed investigation of the Hall effect provides evidence for a prominent Fermi-surface reconstruction as the system is pushed into the high-field state.
\end{abstract}

                         
\maketitle

\section{\label{Introduction}Introduction}
Compounds containing uranium cover a wide range of electronic behavior. Depending on the crystalline surrounding, the uranium $5f$ electrons cause various electronic ground states, e.g., a heavy fermion liquid (UPd$_{2}$Al$_{3}$~\cite{Geibel1993,Jourdan1999}), hidden order (URu$_{2}$Si$_{2}$~\cite{Palstra1985}), Mott insulator (UO$_{2}$~\cite{Gilbertson2014}), insulator (USe$_{3}$~\cite{Shlyk1995}), or even unconventional superconductivity (UPt$_3$~\cite{Stewart1984}, UTe$_{2}$~\cite{Aoki2019}). This diversity is based on the similar energy scales of the $5f$-electron bandwidth, $f-f$ Coulomb, spin-orbit, and exchange interaction. In particular, the $5f$-electron correlation effects are a topical issue for uranium compounds~\cite{Hewson1993,Fujimori2012,Troc2016,Chen2019,Fujimori2019}.

Uranium mononitride (UN) is one example of an intensively studied compound with a rather simple crystal structure (face centered cubic~\cite{Rundle1948, Mueller1958}), but yet not fully understood electronic and magnetic structure~\cite{Atta-fynn,Samsel-Czekaa2007,Fujimori2012,Troc2016,Shrestha2017,Fujimori2019,Soderlind2019}. There are indications for a dual nature of the uranium $5f$ electrons with both local and itinerant character: On one hand, a local origin of antiferromagnetic (AFM) ordering is evidenced by a Curie-Weiss dependence of the magnetic susceptibility above the N\'eel temperature, $T_N \approx 53\,$K, with an effective moment close to the free-ion value. Grunzweig-Genossar \textit{et al.}~\cite{Grunzweig1968} presented a first theoretical model based on localized $5f$ electrons that couple via RKKY interaction. This model fails though to explain the large electronic specific heat at low temperature~\cite{Troc2016}. Recent magnetization, magnetostriction, and ultrasound measurements revealed a magnetic-field-induced spin-flop transition at a critical field of about $H_{SF}=60\,$T, supporting the local picture~\cite{Shrestha2017,Gorbunov2019}. Moreover, the distance of $0.346\,$nm (close to the Hill limit~\cite{Hill1970}) between the uranium atoms in UN suggests a weak orbital overlap and, hence, a significant localization~\cite{Samsel-Czekaa2007}.

On the other hand, the itinerant character of the $5f$ electrons is evidenced by a clear deviation of UN's small ordered magnetic moment, $0.75\mu_B/U$, from the free-ion value and a strong pressure dependence of $T_N$ similar to that of the ordered moment~\cite{Fournier1980}. Further evidence for an itinerant picture was provided by inelastic neutron scattering~\cite{Holden1984}, angle-resolved photoemission spectroscopy (ARPES) measurements~\cite{Fujimori2012} and band-structure calculations~\cite{Atta-fynn,Samsel-Czekaa2007}. Substitution with C on the N site revealed, that the AFM order in UN depends strongly on the next-nearest-neighbor environment~\cite{Lander1974}.

A dual nature of the $5f$ electrons was suggested from earlier ARPES studies~\cite{Ito2001} that discovered a large $5f$ contribution at the Fermi level and the presence of non-dispersive bands. On top of that, nuclear magnetic resonance (NMR) studies of $^{14}$N in UN confirmed the participation of \textit{f} electrons in the formation of both the conduction bands and the localized moments~\cite{Kuznietz1969}.

The type of AFM order and associated lattice distortion at $T_N$ of UN are still debated. The first work by Curry reported a $1k$-type antiferromagnet with the moments along $<100>$ ~\cite{Curry1965}. Subsequently, an x-ray study suggested a small tetragonal distortion, $\left|c/a – 1\right| = 6.5\times10^{-4}$, compatible with the $1k$-type AFM order~\cite{Marples1974}. Knott \textit{et al.} could not confirm the presence of this distortion and concluded that it should be much less than $6.5\times10^{-4}$, if any~\cite{Knott1980}. Finally, a recent study of UN thin films by Bright \textit{et al.} also found no spontaneous distortion despite the strong magnetoelastic interactions~\cite{Bright2019}. This puts into question the widely accepted $1k$ AMF structure of UN. In fact, the absence of any distortion of the cubic crystal structure of UN is compatible with a $3k$ AFM structure that was found, e.g., in UO$_2$~\cite{Blackburn2005,Caciuffo2007,Jaime2017}.
The ordered AFM ground state is very robust against the application of magnetic field. A polarized state is induced via a spin-flop transition just below $60\,$T~\cite{Troc2016}. The induced magnetic moment above the metamagnetic transition reaches with $0.3\,\mu_B/$U only a third of the full effective magnetic moment and suggests only a partial alignment of the spins~\cite{Shrestha2017}. Assuming a layered Ising-spin model the $H-T$ phase diagram was reproduced and complete alignment of the spins was predicted to occur at $258\,$T~\cite{Troc2016}. The metamagnetic transition was so far observed via magnetization~\cite{Troc2016}, dilatometry~\cite{Shrestha2017}, magnetocaloric, and ultrasound measurements~\cite{Gorbunov2019}. Magnetostriction measurements indicate a change in the nature of the transition from first to second order at a tricritical point at $24\,$K and $52\,$T~\cite{Shrestha2017}. Magnetostriction and sound velocity detect a second field-induced anomaly at about $10\,$T, that likely stems from a rearrangement of magnetic domains~\cite{Shrestha2017, Gorbunov2019}.

In order to improve access to the low-temperature and high-field electronic properties of the highly conducting metal UN we use focused ion beam (FIB) assisted micromachining for the fabrication of high-aspect-ratio devices. This approach enables us to measure highly conductive metals in a pulsed-field environment with high precision and to gain insights into fundamental electronic and magnetic properties, and their changes at large fields. Measurements of the Hall effect provide information about changes of the charge-carrier distribution with electronic states at the Fermi level, i.e., changes of the Fermi-surface (FS) topology.

We present results from magnetic torque and magnetotransport measurements in single crystals of UN. We fabricated thin slices as well as micron-sized transport devices. The achieved precision in combination with high pulsed magnetic field and low temperature help us to reveal the special nature of the high-field transition as well as the strong effect on the Hall effect and, hence, the charge carriers at the Fermi level. We confirm the evolution of the transition from a broad and split appearance into a sharp single step as temperature is decreased. Our pulsed-field Hall data provide evidence for a significant FS reconstruction at the high-field transition.

\section{\label{Experimental Details}Experimental Details}

\begin{figure}[tb]
\includegraphics[width=\linewidth]{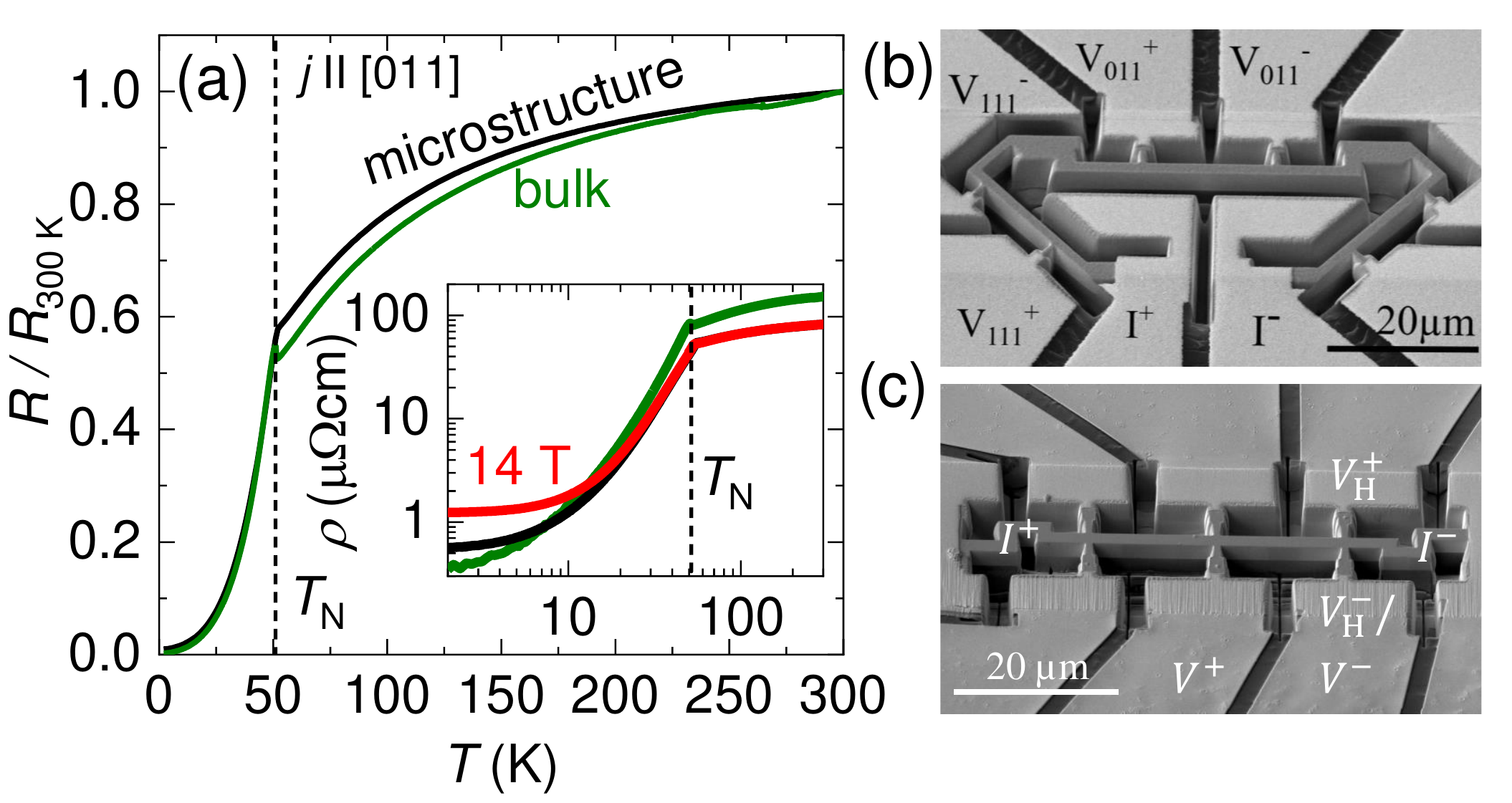}
    \caption{(a) Temperature-dependent resistance of the FIB device A (black) and a bulk sample (green) scaled to the room-temperature value. The inset shows resistivity curves recorded at 0 and $14\,$T. (b),(c) Scanning electron microscope images of the FIB structures A and B with (b) triangular and (c) Hall-bar geometry, respectively (see text).
    }
    \label{R_T}
\end{figure}

The UN single crystals were prepared as described in Ref.~\onlinecite{Troc2016}. We produced several microstructures out of one oriented single crystal which was previously used for ultrasound measurements~\cite{Gorbunov2019}. The micromachining was done with the help of a Ga or Xe focused ion beam. A detailed description of the fabrication process can be found elsewhere~\cite{Moll2018}. This approach enables geometries suitable for high-precision electrical-transport measurements on metallic, i.e., highly conductive materials with current running along any desired direction. Two different devices were produced: a triangular shaped device (A) and a Hall-bar (B) along $[011]$ with cross sections of $(2\times 4)\,\mu$m$^2$ and $(2\times 2)\,\mu$m$^2$, respectively [see Fig.~\ref{R_T}(b) and (c)]. Pulsed high-field experiments were conducted at the Dresden High Magnetic Field Laboratory in a $70\,$T pulsed-magnet system with a pulse duration of 150\,ms equipped with a $^4$He cryostat insert. Static-field measurements were done in a $14\,$T Quantum Design PPMS superconducting-magnet system. Resistivity measurements were performed with a standard a.c. four-point lock-in technique in a \textsuperscript{4}He flow cryostat.

For the torque measurements we used a Wheatstone-bridge-balanced piezo-resistive cantilever (eigenfrequency ~ 300\,kHz)~\cite{Ohmichi2002} loaded with a sample in a $^3$He cryostat insert. The setup was mounted on a rotator, such that the angle between field and cantilever could be varied. Pulsed magnetic fields of up to $68\,$T were applied. We investigated two cuboid samples with different dimensions: a small ($90\times 20\times 10)\,\mu$m$^3$ and a large one $(180 \times 30 \times 50)\,\mu$m$^3$.

\section{\label{Experimental Results}Experimental Results}
\subsection{Zero-field resistivity}

To verify that the microstructuring process does not alter the sample properties we compare the temperature dependence of the zero-field resistivity between bulk and microstructure samples, see Fig.~\ref{R_T}(a). The overall temperature dependence, i.e., a monotonic but weak decrease of $\rho$ followed by a kink at $T_N$ and a sharp drop below it, is not altered. Some details differ though: the residual resistivity ratio (RRR) is reduced from approximately 400 in the bulk sample to 150 in the microstructured device, see inset of Fig.~\ref{R_T}(a). Strain, induced by the substrate, may be responsible for the reduced high-temperature value of $\rho$. A systematic study may be required to fully understand these discrepancies, but goes beyond the scope of this work.

The residual resistivity $\rho_0$ of $0.5\,\mu\Omega$cm at $1.5\,$K is close to the bulk value ($\approx 0.3\,\mu\Omega$cm), proving the high quality of our single crystals. In accordance with previous measurements the resistivity exhibits an exponential dependence for temperatures below $T_N$. Above the transition it traces a monotonically, weakly growing course saturating around room temperature~\cite{VanDoorn1977c}. The anomalous high-temperature dependence of $\rho$ has been associated with a change in the crystal-field splitting of the $5f$-electron levels and additional phonon contributions~\cite{Samsel-Czekaa2007,Pinto1996}.

According to Samsel-Czeka\l{}a \textit{et al.}~\cite{Samsel-Czekaa2007}, the zero-field resistivity in the ordered phase of an AFM metal can be described within the theoretical formalism provided by Andersen and Smith~\cite{Andersen1979} via $\rho(T)=\rho_0 + AT^2+ BT(1+ 2T/\Delta)\exp(-\Delta/T)$, where $A$ and $B$ are constants and $AT^2$ describes a standard Fermi-liquid-like dependence. The last term represents scattering due to AFM order with a spin-density-wave gap estimated to a value of $\Delta\approx 165\,$K for UN~\cite{Samsel-Czekaa2007}. In the inset of Fig.~\ref{R_T}, we show resistivity data recorded in $14\,$T. The exponential temperature dependence is also observed at higher fields, below the field-induced transition, see Fig.~\ref{Resistivity}(a) and (b), and the description in the next section.

\subsection{High-field magnetoresistance}

\begin{figure}[tb]
	\includegraphics[width=\linewidth]{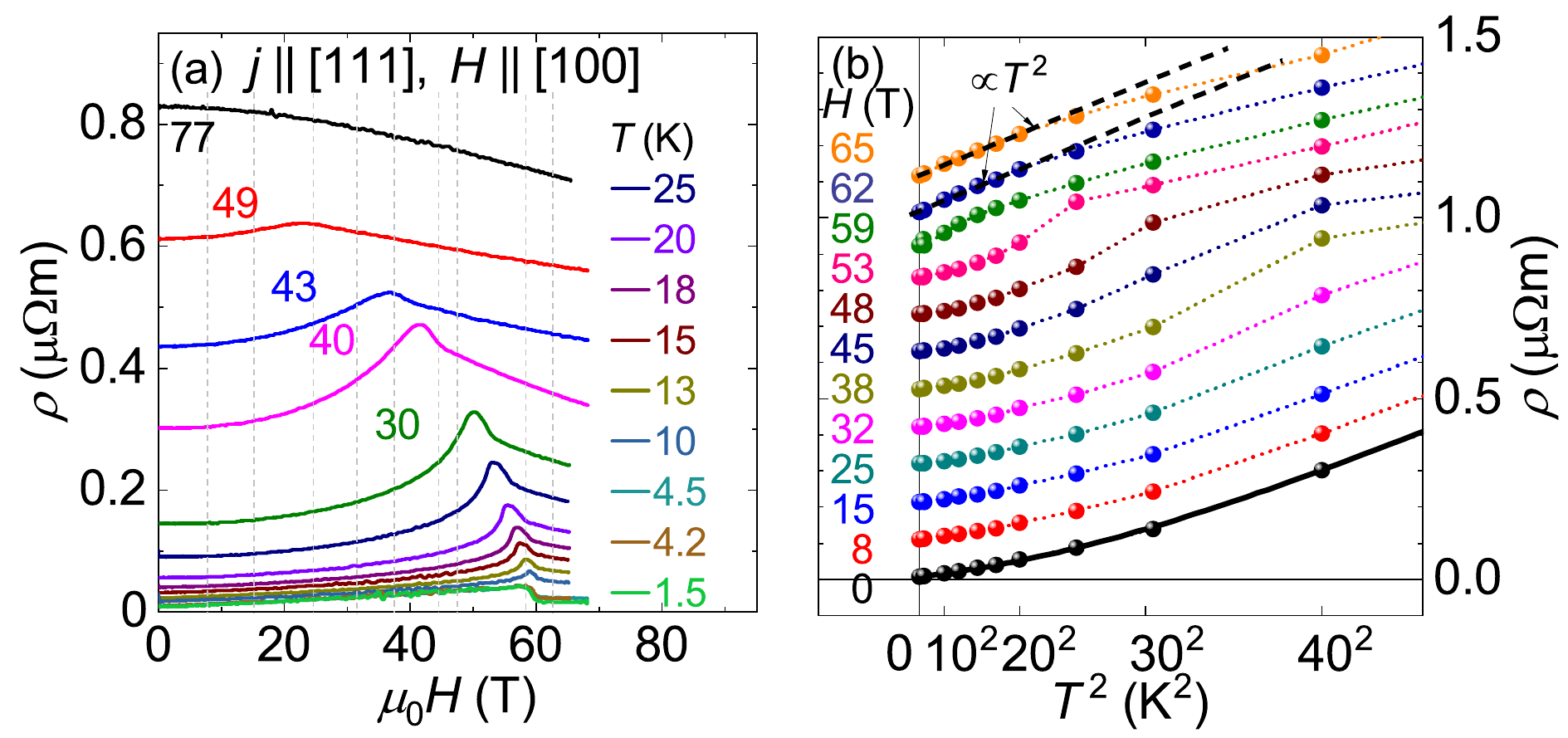}
	\caption{(a) Field-dependent resistivity at various temperatures. The sharp step-like feature for $T<10\,$K turns into a maximum that broadens as we increase temperature. Up and down sweeps match perfectly (no sign of eddy current heating during the pulse). (b) temperature-dependent resistivity extracted from cuts at constant fields, marked by grey dotted lines in (a). While for zero field the resistivity exhibits an exponential dependence (as described in the main text), it is proportional to $T^2$ at low temperature for fields above $60\,$T, as indicated by the black dashed lines for the 62 and $65\,$T data. The curves are shifted consecutively by $0.1\,\mu\Omega$cm for better visibility.}
	\label{Resistivity}
\end{figure}
\begin{figure}[tb]
	\includegraphics[width=0.85\linewidth]{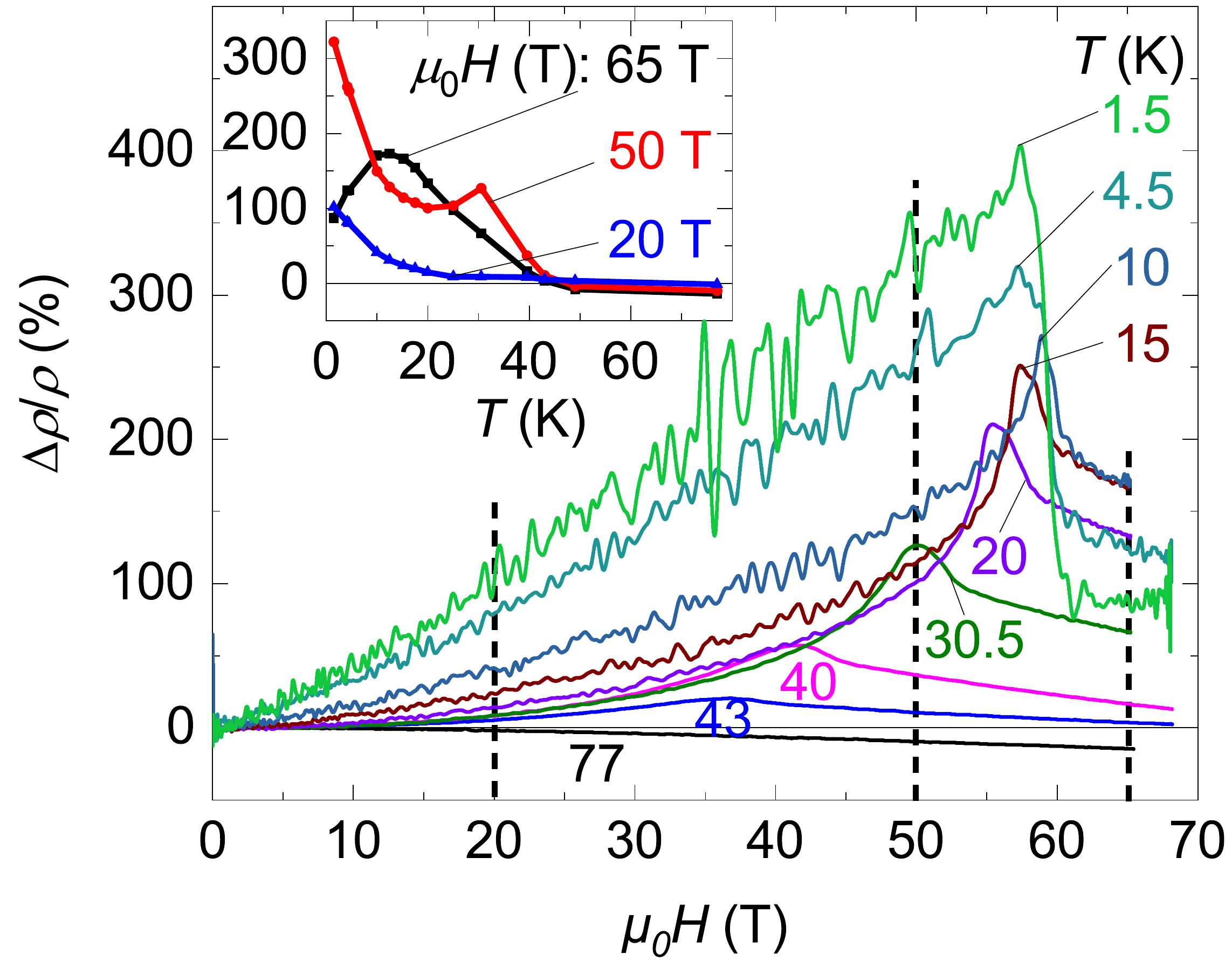}
	\caption{Relative high-field magnetoresistance of UN, sample A, at various temperature values extracted from Fig.~\ref{Resistivity}(a). The inset shows $\Delta\rho/\rho(T)$ at fixed field values, marked by the vertical dashed lines.}
	\label{MR}
\end{figure}
UN experiences a spin-flop transition just below $60\,$T for field applied along any of the principal crystallographic directions~\cite{Gorbunov2019}. For external magnetic field applied along $[100]$ we observe a sharp drop in the resistivity at $H_{SF}\approx60\,$T for the data recorded with $T< 10\,$K [see Fig.~\ref{Resistivity}(a)]. For  $10\,$K$\,\leq T < T_N$ the overall shape of the transition changes - we observe a pronounced broad maximum at the transition followed by a change in slope.

In order to trace the temperature dependence of the resistivity for high fields we extract values at constant field from the isothermal pulse-field data in Fig.~\ref{Resistivity}(a) and plot them versus $T^2$ in Fig.~\ref{Resistivity}(b). The exponential temperature dependence is traceable for low fields below the critical high-field transition at $H_{SF}$, discernible by the change in slope moving to lower temperature values for increasing field [Fig.~\ref{Resistivity}(b)]. Above $H_{SF}$, the resistivity is proportional to $T^2$ for $T\leq20\,$K, highlighted by the linear fits (black dashed lines) in Fig.~\ref{Resistivity}(b) for the 62 and $65\,$T data. Accordingly, the high-field state exhibits a simple metallic character, as a squared temperature dependence is the hallmark of a Fermi liquid. The deviation at higher temperatures may be explained by increasing phonon scattering~\cite{Samsel-Czekaa2007}. We did not observe quantum oscillations in the accessible experimental temperature and field range, which may be related to heavy effective cyclotron masses of the charge carriers.

In Fig.~\ref{MR}, we provide the relative magnetoresistance (MR) $\Delta\rho/\rho=(\frac{\rho(H)}{\rho(H=0)}-1)\times100\,\%$. In this representation, the evolution of the high-field transition from a broad peak-like appearance into a single step at low temperature is more apparent. For $T<10\,$K, the MR rises almost linearly with field until it experiences a sharp drop after which its slope turns negative. Above $10\,$K, the MR exhibits an almost exponential growth and culminates in a hump-like feature, associated with the high-field transition that broadens and shifts to lower fields as we further increase temperature. The temperature-dependence of MR is shown in the inset of Fig.~\ref{MR}. Above the critical field, the slope of MR turns positive at lowest temperature with a maximum between 10 and $15\,$K.
\begin{figure}[tb]
	\includegraphics[width=0.95\linewidth]{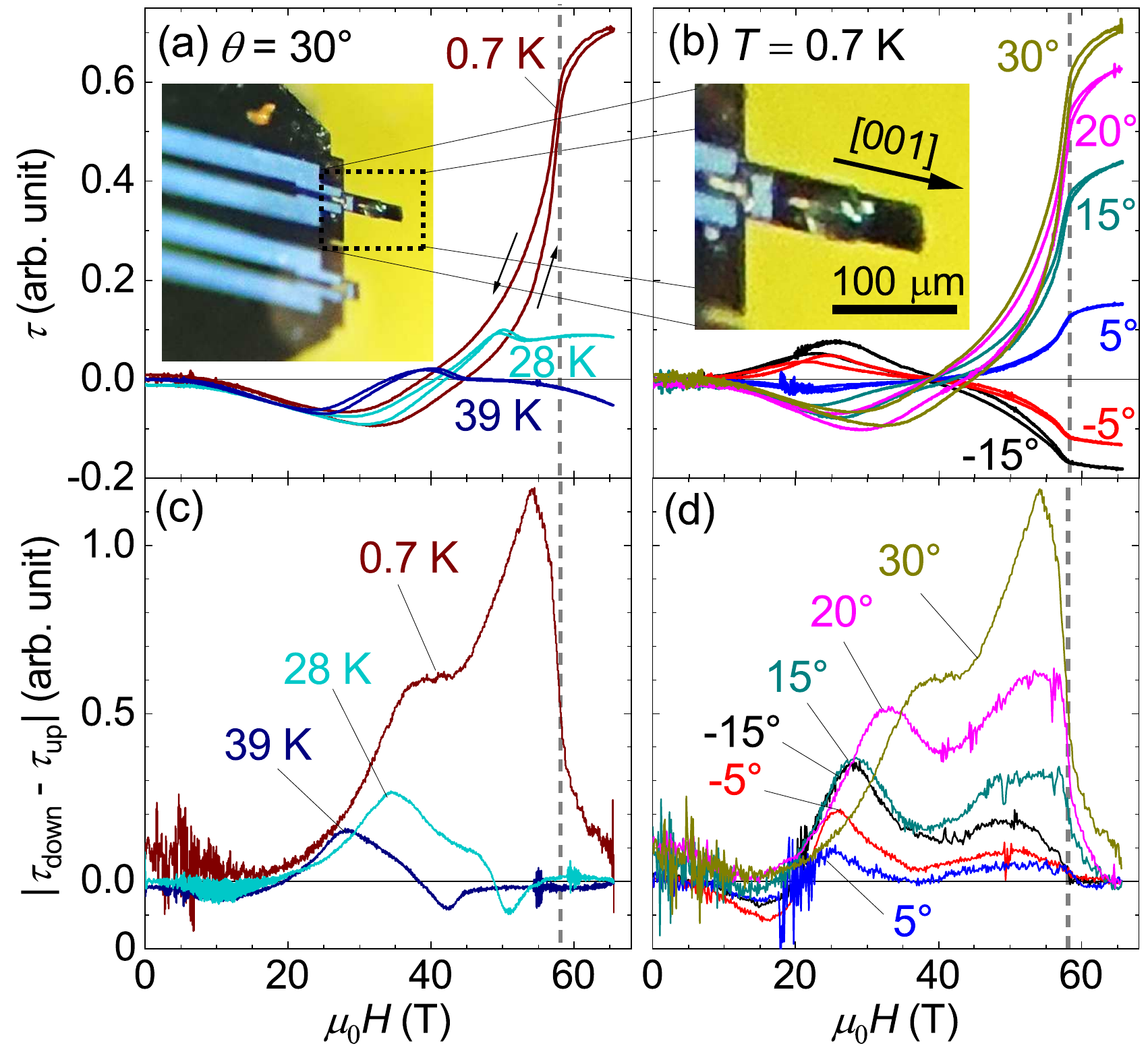}
	\caption{(a) Field-dependent magnetic-torque data recorded for constant tilt angle of $\theta=30^\circ$ off $[100]$ at different temperature values and (b) for constant $T=0.7\,$K at different tilt angles. The insets show an images of the crystal attached to a $120\,\mu$m long piezo-resistive microcantilever. (c), (d) Difference between up and down field sweeps of the magnetic-torque data in (a) and (b). The vertical dashed lines mark the critical field $H_{SF}$ at base temperature.}
	\label{Torque}
\end{figure}

\subsection{Magnetic torque}

Let us now turn to our results from pulsed-field measurements of magnetic torque presented in Figs.~\ref{Torque}(a) and (b). The sample investigated was a thin platelet of UN with thickness similar to the one of the transport microstructures. The insets of Fig.~\ref{Torque}(a) and (b) show an images of the piezo-resistive cantilever including the sample attached by silicone grease. In Fig.~\ref{Torque}(a) we present isothermal curves recorded at constant tilt angle, $\theta=30^\circ$ off the $[100]$ direction towards $[010]$. A similar set of angle-dependent torque data for a second sample with 10 times smaller volume can be found in the appendix (Fig.~\ref{Torque2}). For the elevated temperatures $T= 28\,$K and $39\,$K, we observe a broad maximum (surrounded by two minima) that resembles the behavior we observe in the resistivity near the transition field. In Fig.~\ref{Torque}(b) we show field-dependent torque data at various angles and at base temperature of $0.7\,$K. A broad extremum is observable (maximum/minimum depending on the angle) at a field between $20$ to $35\,$T. There, the change in magnetization is maximum, which affects the magnetic anisotropy sensed by the torque. The amplitude of the extremum depends on the sample size, i.e., for a $10\times$ smaller lamella it is much weaker (see Appendix Fig.~\ref{Torque2}). The extremum is followed by a strong increase of the absolute torque signal. Since the cantilever is not equally sensitive for the two bending directions the curves are not symmetric about zero deflection angle. The field value of the step-like transition at high field hardly varies with angle.

All curves exhibit a hysteresis between the up and down field sweep that starts to close once the critical transition field is reached. The difference between each up and down sweep, $\Delta\tau=\left|\tau_\mathrm{down}-\tau_\mathrm{up}\right|$ is presented in Fig.~\ref{Torque}(c) and (d). $\Delta\tau$ is most pronounced for larger tilts away from the principal axis and exhibits two distinct maxima. As we increase temperature, the lower peak shifts towards smaller fields. The same trend occurs for a reduction in the tilt angle at constant temperature, shown in Fig.~\ref{Torque}(d). However, the maximum at higher field subsides while exhibiting no apparent angle dependence.  

\subsection{Hall effect}

\begin{figure}[tb]
	\includegraphics[width=0.95\linewidth]{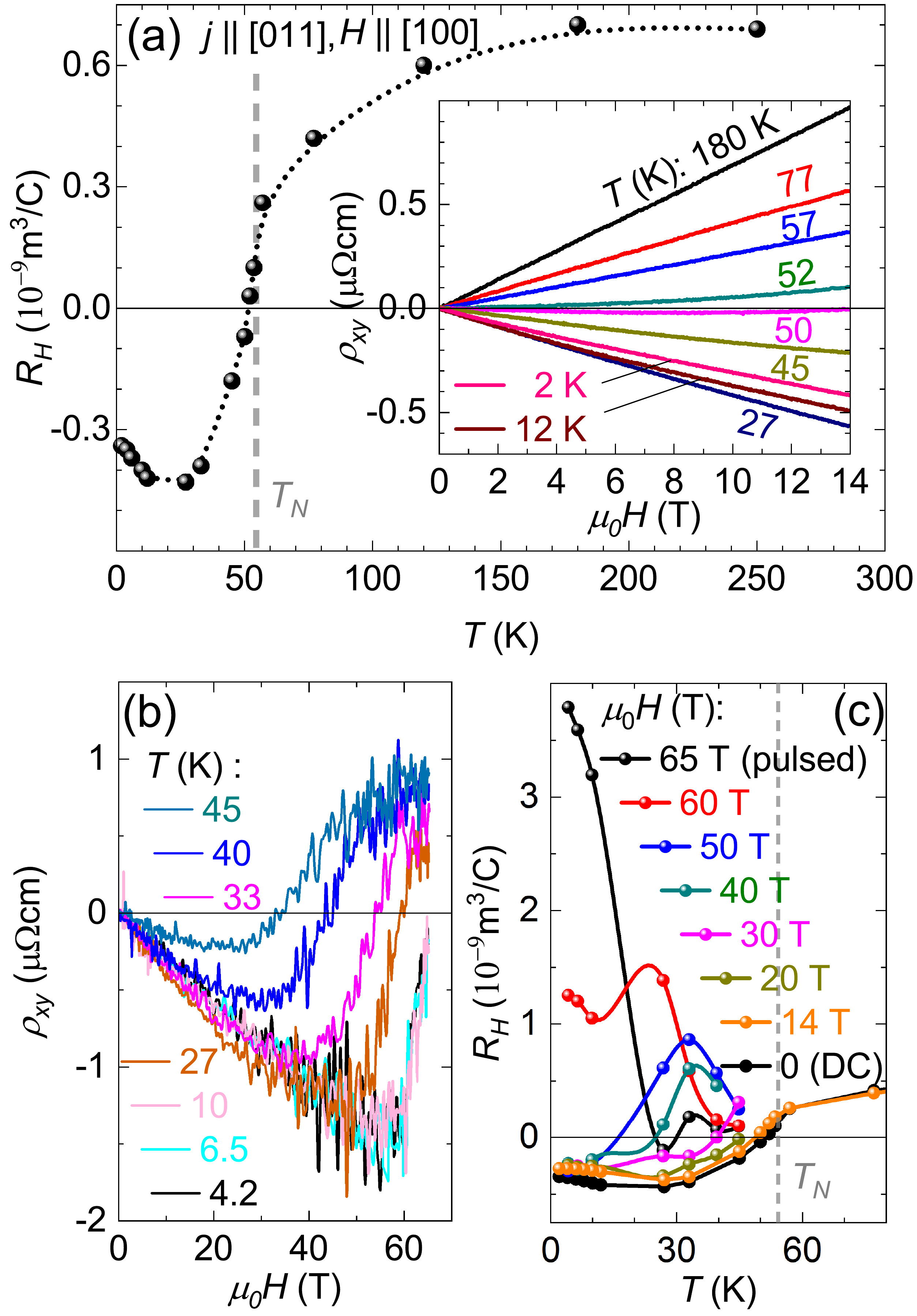}
	\caption{(a) Hall coefficient determined from the low-field ($\mu_0H\leq4\,$T) slope of the Hall-resistivity data shown in the inset recorded in steady fields. (b) Hall resistivity recorded in pulsed magnetic fields at various temperatures. (c) Hall coefficient, $R_H$, extracted from first derivatives (shown in Fig.~\ref{Hall_derivatives}) of the data in (b) at fixed field values.
	}
	\label{Hall}
\end{figure}

Next, we will discuss results from Hall measurements conducted for sample B. Figure~\ref{Hall}(a) presents the temperature dependence of the low-field Hall coefficient, $R_H$, extracted from the slope of the Hall resistivity, $\rho_{xy}(H)$, between zero and $4\,$T. The inset of Fig.~\ref{Hall}(a) shows isothermal field sweeps recorded in DC fields. $R_H$ varies between about $0.7\times10^{-9}\,$m$^3/$C and $-0.4\times10^{-9}\,$m$^3/$C corresponding to $9\times 10^{29}$m$^{-3}$ hole- and $16\times 10^{29}$m$^{-3}$ electron-like carriers in a single-band approximation, respectively. Entering the ordered phase leads to a sign change in the low-field Hall resistivity: For temperatures above $T_N$, $\rho_{xy}$ exhibits a positive, almost linear slope; It turns negative as soon as the temperature falls below $T_N$. This dependence is similar to what was reported for USb, where AFM order sets in at a temperature of about $215\,$K~\cite{Schoenes1984}.

There are two key features in the temperature dependence of the low-field Hall resistivity: The first one occurs at $T_N$, where $\rho_{xy}(T)$ changes its sign, and the second one between $27$ and $12\,$K, where $R_H$ reaches its minimum in Fig.~\ref{Hall}(a). This slope change below $27\,$K is also discernible in the low-field part of the Hall data obtained in pulsed fields [see Fig.~\ref{Hall}(b)]. Furthermore, for $T\leq 10\,$K the pulsed-field Hall-resistivity curves lie on top of each other within the signal-to-noise limited resolution. Above this temperature range they start deviating, see Fig.~\ref{Hall}(b). The field-dependent Hall-effect curves display an S-shape as the signal changes from negative with negative slope towards positive with positive slope, going from low to high magnetic field, respectively. Therefore, a strong maximum shows up in the high-field Hall coefficient, shown in Fig.~\ref{Hall}(c) [The respective first derivatives can be found in the Appendix, Fig.~\ref{Hall_derivatives}].

Interestingly similar to the magnetoresistivity, the high-field transition in $\rho_{xy}(H)$ sharpens in the low-temperature range, for $T=10\,$K and below. This resembles the tricritical behavior previously reported at $24\,$K and $52\,$T~\cite{Shrestha2017}. In order to learn more about the slope of the Hall resistivity at high field and low temperature, larger fields above $70\,$T are necessary. This however, is beyond the value accessible in this work.

\begin{figure}[tb]
	\includegraphics[width=\linewidth]{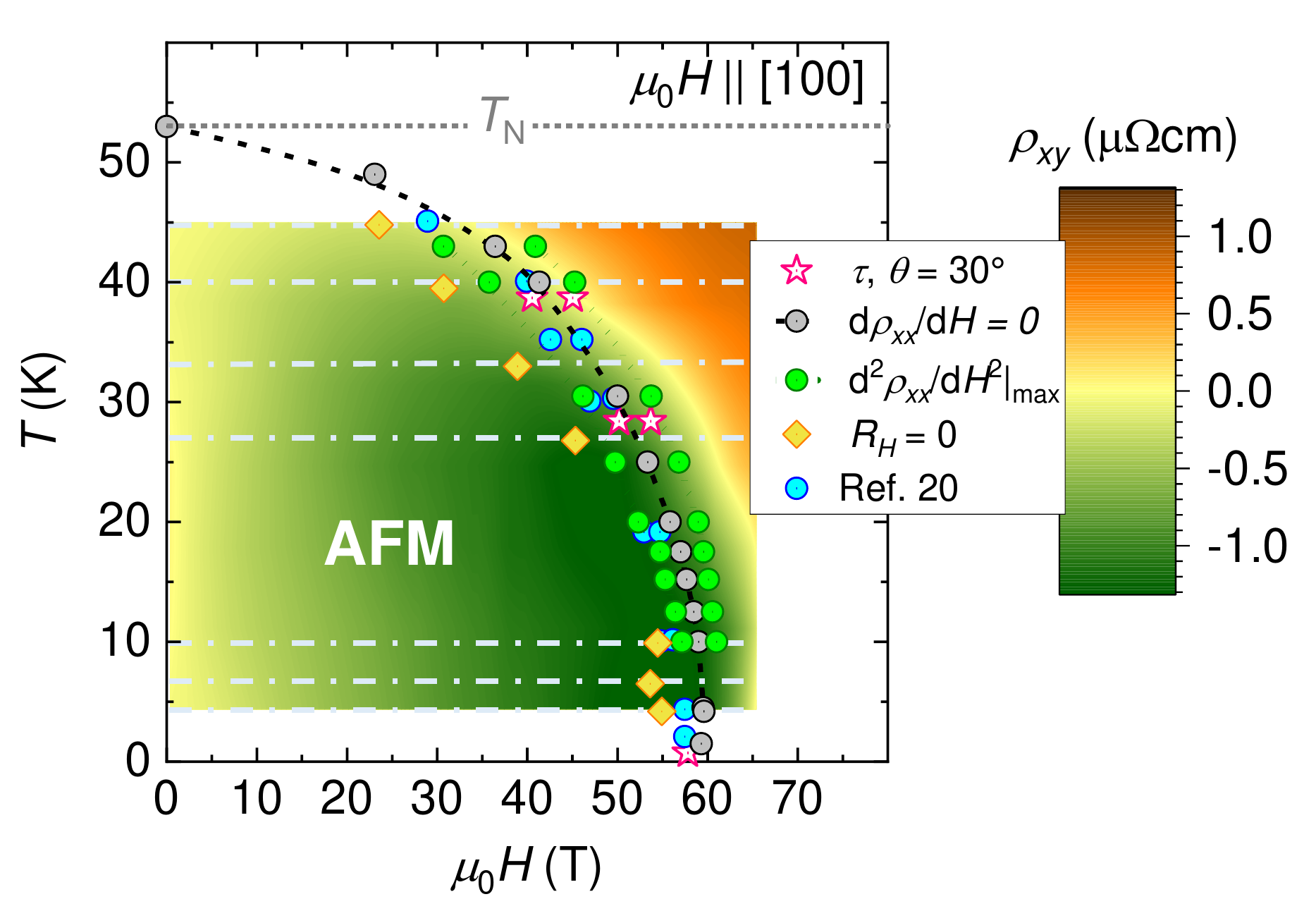}
	\caption{Schematic $H - T$ phase diagram of UN for $\mu_\mathrm{0}H||[100]$. Data points mark the respective features, related to the high-field transition, observed in magnetic torque (magenta stars), magnetoresistivity (grey and green circles), and Hall effect (orange diamonds) in comparison to transition fields reported from ultrasound (blue circles) by Gorbunov \textit{et al.}~\cite{Gorbunov2019}. First and second derivatives are shown in the appendix. The high-field phase transition appears more complex at elevated temperatures. The dashed line is a guide to the eye. Error bars are of the same order as the symbol sizes. The color map presents the interpolated pulsed-field Hall data recorded at constant temperatures (highlighted by light grey dash-dotted lines) from Fig.~\ref{Hall}(b).
	}
	\label{PD}
\end{figure}

We summarize our results from resistivity, torque, and Hall measurements in the phase diagram in Fig.~\ref{PD}. It is in accordance with previous ultrasound and magnetization studies~\cite{Gorbunov2019} (see blue circles). The prominent magnetic phase transition, associated with a spin flip, occurs at approximately $58\,$T applied parallel to the $[100]$ direction for $T=1.5\,$K. The step-like appearance in resistivity and torque resembles a transition of first order. At first glance, $H_{SF}$ decreases monotonically to zero upon approaching $T_N$. However, above $10\,$K, we observe a clear transformation of the transitional behavior: The initially sharp step transforms into a broad hump followed by a change in slope, which is reflected by multiple extrema in the second derivatives (green dots in Fig.~\ref{PD}). The respective first and second derivative data can be found in the Appendix, Fig.~\ref{Resistivity_derivatives}. The color map presents an interpolation of the pulsed-field Hall data shown in Fig.~\ref{Hall}(b). $\rho_{xy}(H)$ changes sign as the samples transitions into the high-field phase. The orange diamonds mark the field values where $R_H=0$.  

\section{\label{Discussion}Discussion}

The very low resistivity ($\sim 0.5\,\mu\Omega$cm) of UN poses a challenge to transport experiments at high magnetic fields. Here, our approach of FIB microstructuring offers a clear advantage, as it helps boost the signal-to-noise ratio. Since the temperature-dependent resistivity curves of bulk and micromachined samples exhibit the same behavior, we can validate our approach of microstructuring [Fig.~\ref{R_T}(a)].

The magnetic ground state of UN is yet to be determined unambiguously. Our results provide further evidence for the unusual nature of the field-induced critical transition at elevated temperatures: The comparison of magnetoresistivity and magnetic torque reveals a broad feature in the vicinity of the high-field transition that coincides with split transition line reported previously~\cite{Gorbunov2019}. The monotonic increase in resistivity followed by the broad hump signals a significant field-dependent change in the main scattering channels [Fig.~\ref{Resistivity}(a)]. The monotonic, nearly quadratic slope of the relative magnetoresistance changes to an almost linear field-dependence below $10\,$K (see Fig.~\ref{MR}) and a single step-like transition emerges. This is further evidence for a drastic change in the magnetic scattering within the ordered phase. A temperature- and field-dependent reduction of the resistivity below $T_N$ and above the critical transition-field is typical for AFM transitions~\cite{Rapp1978,Schoenes1984,Helm2020}, associated with a reconstruction of the band structure in the paramagnetic or field-polarized state. At high fields, above the transition, the resistivity exhibits a quadratic temperature dependence, i.e., metallic Fermi-liquid-like conduction, at least at low temperatures. Hence, the AFM scattering is fully suppressed and a metallic state is achieved.

Furthermore, the low-field Hall coefficient changes its sign from positive to negative at $T_N$, providing clear evidence for an event that affects the states near the Fermi level. Similar behavior was also observed for USb, where the zero-field magnetic order was confirmed to be of $3k$ type~\cite{Magnani2010}. Band-structure calculations predict a complex FS with multiple bands at the Fermi level~\cite{Troc2016,Samsel-Czekaa2007}, failing, however, to incorporate magnetism from the $5f$ electrons. Once AFM order establishes, the new magnetic Brillouin zone crosses parts of the FS at certain points in $k$ space, leading to a reconstructed FS that obeys the Luttinger theorem~\cite{Luttinger1960}. Recent ARPES experiments observed indications for such a scenario~\cite{Fujimori2012}. A redistribution of electron- and hole-like parts of the FS due to a superpotential associated with density-wave order has been proposed as origin for small FS sheets observed in cuprate superconductors~\cite{Helm2009,Doiron-Leyraud2007}. In these compounds, a field-dependent sign change of $R_H(H)$, from negative to positive, at high field supports such an reconstruction scenario. For UN, the gradual change of $R_H$, which starts deviating from its linear field-dependence far below the critical AFM suppression field, resembles results reported from transport measurements in AFM 2D conductors~\cite{McKenzie1996,Helm2015}. At high enough magnetic fields the unreconstructed FS is recovered by the so-called magnetic-breakdown effect~\cite{Falicov1965}. Alternatively, the broad and enhanced Hall response at elevated temperatures below $T_N$ could originate from critical magnetic fluctuations~\cite{Niu2020,Aoki2019a} or domain formations. The latter may also cause the observed hysteretic response in the magnetic torque and was recently suggested from magnetoelastic high-field investigations~\cite{Gorbunov2019}.

At low temperature, the transition in the Hall resistivity becomes narrower and temperature independent. This matches the sharp step-like behavior observed in the longitudinal resistivity and magnetic torque. Moreover, we find that the high-field MR exhibits a positive slope in the low-temperature regime (see $65\,$T data presented in the inset of Fig.~\ref{MR}), which indicates a change in the field-induced scattering response. Hence, this evolution supports the idea of a different magnetic state at low temperatures above $H_{SF}$, in line with the reduced magnetic moment and theoretical predictions~\cite{Troc2016}. The $T^2$ dependence in the high-field resistivity indicate a strongly reduced coupling of the charge carriers to the magnetic system. The observed change with decreasing temperature from a second- to first-order-like appearance of the high-field transition resembles similar behavior observed for the AFM heavy fermion CeRh$_2$Si$_2$~\cite{Knafo2017}. Indeed, such a behavior is more common in itinerant FM compounds~\cite{Belitz1999,Brando2016}, but may also be related to a tricritical point~\cite{Shrestha2017}.

\section{\label{Conclusion}Conclusion}

	Our work supports the picture based on a dual nature of the uranium $5f$ electrons in UN. We provide evidence for a sign change in the Hall response, an indication for a FS reconstruction at the critical field linked to the Néel temperature $T_N$. We confirm the peculiar evolution with decreasing temperature of the high-field transition from a more complex and broad phase boundary into a sharp first-order transition at temperatures below $10\,$K. The observed temperature range, however, is significantly lower than the previously reported tricritical point near $25\,$K~\cite{Shrestha2017}. Beyond the spin-flop transition, for $H\geq H_{SF}$, the MR changes its slope, magnetic torque seems to exhibit no hysteresis anymore, and the resistivity exhibits a $T^2$ dependence. This clearly indicates the reduced coupling of any remaining magnetism in the low-temperature high-field state, whose structure is yet to be clarified. To further understand the full band structure of UN, the detection of magnetic quantum oscillations is highly desirable. This may be achievable at temperatures lower than the range accessible in our study. The nature of the magnetic ground state at zero field as well as at high field beyond the spin-flop transition remains an open question. The sign change in the high-field Hall coefficient from negative to positive indicates the suppression of the magnetic order. Nevertheless, the induced magnetic moment was found to be only $1/3$ of the zero field moment. Therefore, both zero- and high-field states call for further experiments that may reveal the magnetic structure in this compound.

\section{\label{Acknowledgments}Acknowledgements}
We would like to thank I. Sheikin and R. Ramazashvili for fruitful discussions. We acknowledge the support of the HLD at HZDR, member of the European Magnetic Field Laboratory (EMFL), and the Deutsche Forschungsgemeinschaft (DFG) through the W\"urzburg-Dresden Cluster of Excellence on Complexity and Topology in Quantum Matter–\textit{ct.qmat} (EXC 2147, Project No. 390858490).
\appendix
\section{Appendix}
\subsection{\label{SuppB} First and second derivatives of the high-field magnetoresistance data}

\begin{figure}[tb]
	\includegraphics[width=0.85\linewidth]{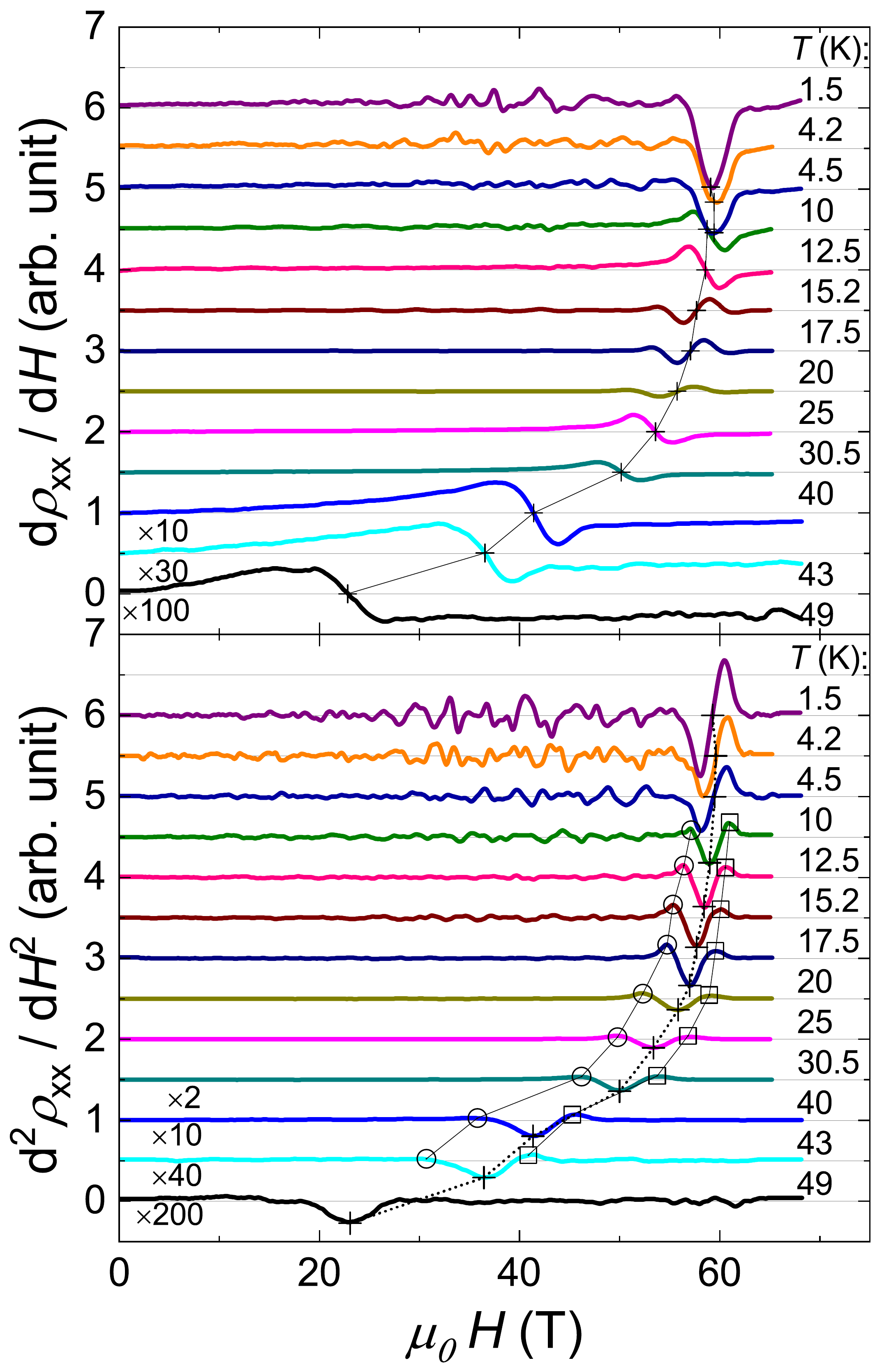}
	\caption{First and second derivatives of the high-field resistivity data shown in Fig.~\ref{Resistivity}(a).
	}
	\label{Resistivity_derivatives}
\end{figure}
In order to extract the transition fields at different temperatures from the raw data shown in Fig.~\ref{Resistivity}, we analyzed the first and second derivatives of the magnetoresistance (see Fig.~\ref{Resistivity_derivatives}). In both representations it immediately becomes clear that the overall appearance of the high-field transition changes significantly below approximately $10\,$K.

\subsection{\label{SuppC}High-field magnetic torque of a second UN sample}

\begin{figure}[tb]
	\includegraphics[width=0.9\linewidth]{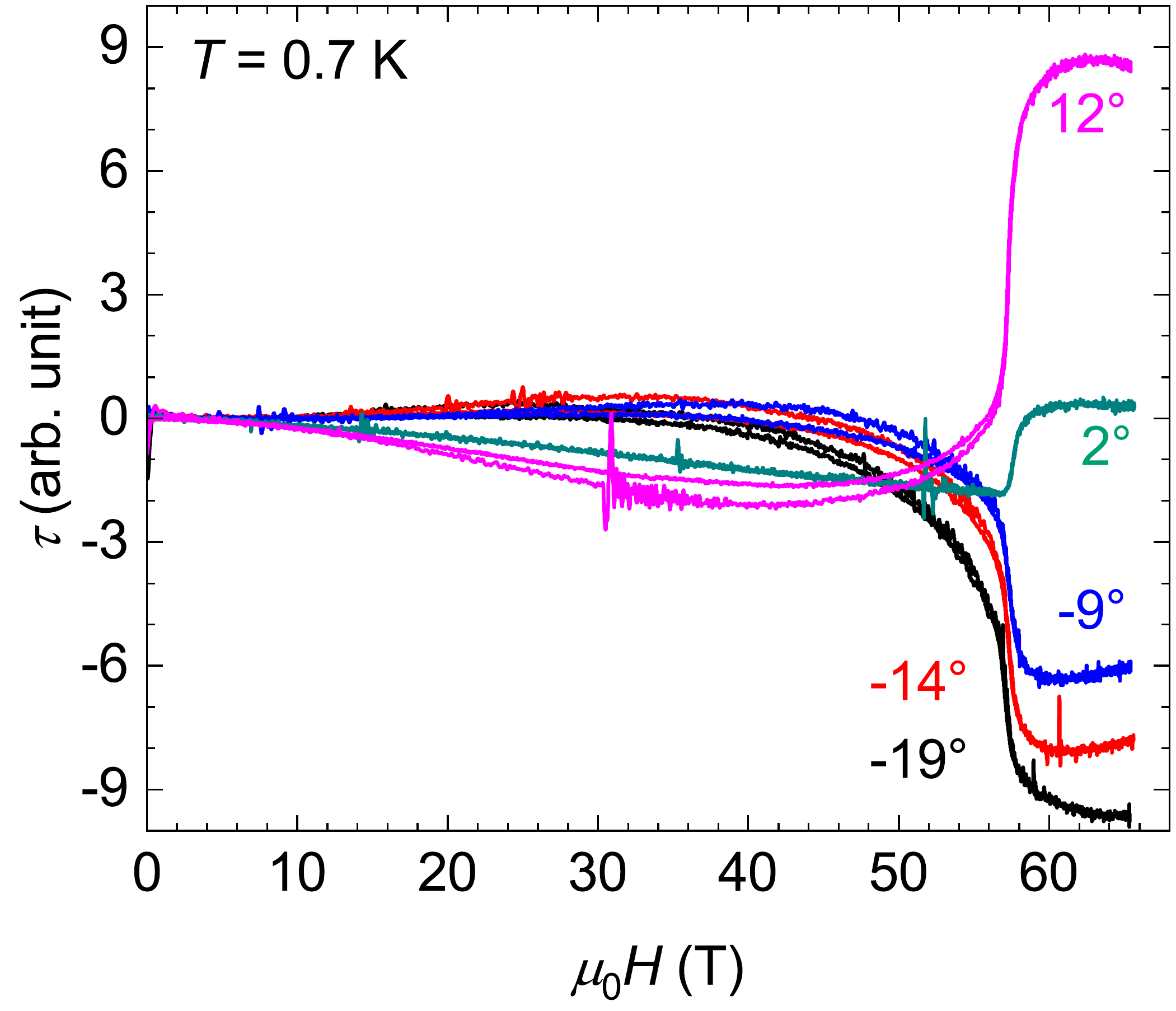}
	\caption{High-field magnetic torque of a second UN sample (with ten times smaller volume) recorded for various tilt angles at $T=0.7\,$K using a piezo-resistive microcantilever.
	}
	\label{Torque2}
\end{figure}
We conducted measurements of the magnetic torque in pulsed fields on a second sample, approximately 10~times smaller in volume as compared to the sample presented in Fig.~\ref{Torque} of the main text. The results are shown in Fig.~\ref{Torque2}. We resolve the metamagnetic transition with a strong, step-like increase (decrease) of the torque signal. Furthermore, the hysteresis between up and down sweep in field and a broad hump between 30 and $45\,$T are reproduced, although not as pronounced as for the larger sample.

\subsection{\label{SuppD} First derivatives of the high-field Hall effect}

\begin{figure}[tb]
	\includegraphics[width=0.85\linewidth]{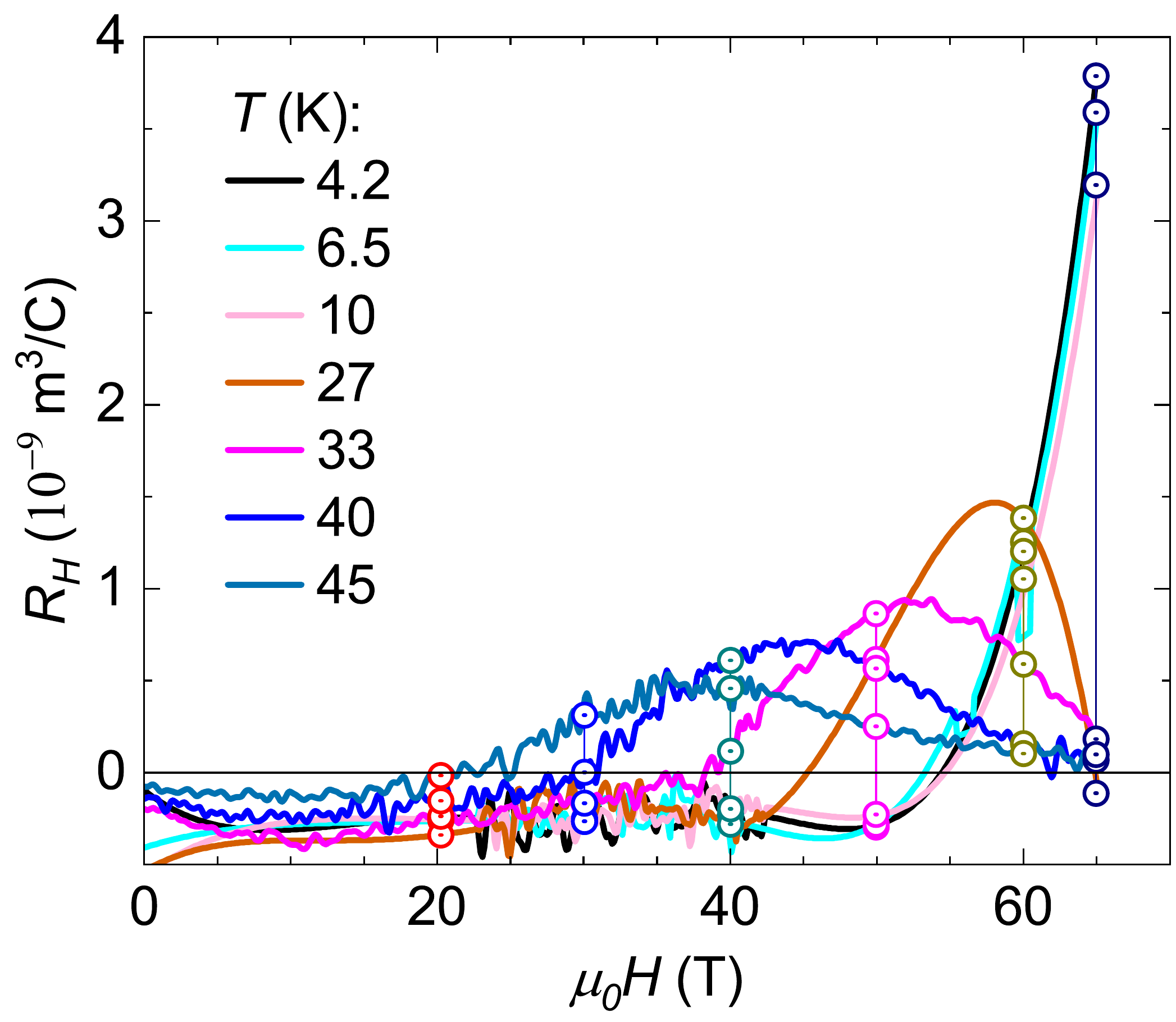}
	\caption{First derivatives of the Hall-resistivity data shown in Fig.~\ref{Hall}(b). Circles are the values at fixed field presented in Fig.~\ref{Hall}(c).
	}
	\label{Hall_derivatives}
\end{figure}
In order to produce Fig.~\ref{Hall}(c), we smoothed the Hall-resistivity data of Fig.~\ref{Hall}(b) by a cubic-spline interpolation. We then extracted the field-dependent Hall coefficient $R_H$ from the first derivative in combination with an additional smoothing step (see Fig.~\ref{Hall_derivatives}).

\newpage
\section{References}
\bibliographystyle{apsrev4-1}
\bibliography{library2}% Produces the bibliography via BibTeX.

\end{document}